\documentstyle[12pt]{article}

\newcommand{\doublespace}{\renewcommand{\baselinestretch}{1.75}
\Large\normalsize}

\def\fff{\displaystyle\frac}
\def\sen{\;\mbox{sen}\,}

\begin{document}
\begin{titlepage}
\title{Energy and angular momentum of the gravitational
field in the teleparallel geometry}
  
\author{J. W. Maluf$\,^{*}$, J. F. da Rocha-Neto$\,^{\S}$  \\
T. M. L. Tor\'{\i}bio,  and K. H. Castello-Branco \\
Instituto de F\'{\i}sica, \\
Universidade de Bras\'{\i}lia\\
C. P. 04385 \\
70.919-970 Bras\'{\i}lia DF\\
Brazil\\}
\date{}
\maketitle

\begin{abstract}
The Hamiltonian formulation of the teleparallel equivalent of
general relativity is considered. Definitions of energy,
momentum and angular momentum of the gravitational field arise
from the integral form of the constraint equations of the
theory. In particular, the gravitational energy-momentum is
given by the integral of scalar densities over a
three-dimensional spacelike hypersurface. The definition for
the gravitational energy is investigated in the context of the
Kerr black hole. In the evaluation of the energy contained
within the external event horizon of the Kerr black hole we
obtain a value strikingly close to the irreducible mass of the
latter. The gravitational angular momentum is evaluated for
the gravitational field of a thin, slowly rotating mass shell.
\end{abstract}
\thispagestyle{empty}
\vfill
\par
\bigskip
\noindent (*) e-mail: wadih@fis.unb.br\par
\noindent (\S) present address: Instituto de F\'{\i}sica
Te\'orica, Universidade Estadual Paulista,  Rua Pamplona 145,
01405-900, S\~ao Paulo, SP, Brazil \par

\end{titlepage}

\newpage
\doublespace
\noindent {\bf I. Introduction}\par

\bigskip
\noindent Teleparallel
theories of gravity have been considered
long time ago in connection with attempts to define the
energy of the gravitational field\cite{Mol}. By studying the
properties of solutions of Einstein's equations that describe
the gravitational field of isolated material systems, it is
concluded  that a consistent expression for 
the energy {\it density} of the gravitational field
would be given in terms of 
second order derivatives of the metric tensor.
It is known that there exists no covariant, nontrivial expression
constructed out of the metric tensor, both in three and in four
dimensions, that contain such derivatives. However, covariant
expressions that contain second order derivatives of tetrad
fields are feasible.  Thus it is
legitimate to conjecture that the difficulties regarding
the problem of defining the gravitational energy-momentum is
related to the geometrical description of the gravitational field,
rather than being an intrinsic drawback of the
theory\cite{Grishchuk}.

It is usually asserted in the literature that the principle of
equivalence prevents the localizability of the gravitational
energy. However, an expression for the gravitational field
energy has been pursued since the early days of general
relativity. A considerable amount of effort has been devoted to
finding viable expressions other than pseudotensors (more
recently the idea of quasi-local energy, i.e., energy
associated to a closed spacelike two-surface, in the context
of the Hilbert-Einstein action integral, has emerged as
a tentative description of the gravitational
energy\cite{Brown}). The search
for a consistent expression for the gravitational energy
is undoubtedly a longstanding problem in general relativity.
The argument based on the principle of equivalence
regarding the nonlocalizability of the
gravitational energy is controversial and not generally
accepted\cite{Grishchuk}. The
principle of equivalence does not preclude the existence of
scalar densities on the space-time manifold, constructed
out of tetrad (or triad) fields,  that may eventually yield the
correct description of the energy properties of the
gravitational field. Such densities may be given in terms of
the torsion tensor, which cannot be made to vanish at a point
by a coordinate
transformation. M\o ller\cite{Mol} was probably the first one
to notice that the tetrad description of the gravitational
field allows a more satisfactory treatment of the gravitational
energy-momentum. 

The dynamics of the gravitational field can be described
in the context of the teleparallel
geometry, where the basic geometrical entity is the tetrad
field $e^a\,_\mu$, ($a$ and $\mu$ are SO(3,1) and
space-time indices, respectively). Teleparallel theories of
gravity are defined on the
Weitzenb\"ock space-time\cite{Weit}, endowed with the affine
connection

$$\Gamma^\lambda_{\mu\nu}=e^{a\lambda} \partial_\mu
e_{a\nu}\;.\eqno(1.1)$$

\noindent The curvature tensor constructed out of Eq. (1.1)
vanishes identically. This connection defines a space-time with
teleparallelism, or absolute parallelism\cite{Schouten}.
This geometrical framework was considered by
Einstein\cite{Einstein} in his attempt at unifying gravity and
electromagnetism. 

Gravity theories in this geometrical framework
are constructed out of the torsion tensor. An infinity of such
theories defined by a Lagrangian density,
quadratic in the torsion tensor, has been investigated 
by Hayashi and Shirafuji\cite{Hay} (who denote $e^a\,_\mu$ as
{\it parallel vector fields}). Among such infinity of theories
a particular one is distinguished, because the tetrad fields that
are solutions of this particular theory yield a metric tensor
that is a solution of Einstein's equations. 
The teleparallel equivalent of general relativity
(TEGR)\cite{Hehl,Kop,Muller,Nester1,Nester2,Maluf1,Per}
constitutes an alternative geometrical description of Einstein's
equations.

A simple expression for the gravitational energy arises in
the Hamiltonian formulation of the TEGR\cite{Maluf1} in
the framework of Schwinger's time gauge
condition\cite{Schwinger}. The energy density is given by a
scalar density in the form of a total divergence that
appears in the Hamiltonian constraint of the
theory\cite{Maluf2}. The investigations carried out so far
confirm the consistency and relevance of this energy expression.

A recent approach to the localization of the gravitational
energy has been considered in the Lagrangian framework of the
TEGR by Andrade, Guillen and Pereira\cite{Per2}. It has
been shown in the latter reference the
existence of an expression for the gravitational energy density
that is a true space-time tensor, and that reduces to
M\o ller's energy-momentum density of the gravitational field.

The Hamiltonian formulation of the TEGR, with no {\it a priori}
restriction on the tetrad fields, has recently been
established\cite{Maluf3}. Its canonical structure is different
from that obtained in Ref. \cite{Maluf1}, since it is
not given in the standard ADM form\cite{ADM}. In this
framework we again arrive at an expression for the gravitational
energy, in strict similarity with the procedure adopted in
Ref. \cite{Maluf2}, namely, by interpreting the Hamiltonian
constraint equation as an energy equation for the gravitational
field. Likewise, the gravitational momentum can be defined. 
The constraint algebra of the theory suggests that
certain momentum components are related to the gravitational
angular momentum. It turns out to be possible to define, in
this context, the angular momentum of the gravitational field.

In this article we investigate the definition of gravitational
energy that arises in Ref. \cite{Maluf3}, in the framework
of the Kerr metric tensor\cite{Kerr}. The whole formulation
developed in  Ref. \cite{Maluf3} is carried out without
enforcing the time gauge condition. It turns out, however, that
consistent values for the gravitational energy are achieved by
requiring the tetrad field to satisfy ({\it a posteriori}) the
time gauge condition. 

We investigate 
the irreducible mass $M_{irr}$ of the Kerr black hole.
It is the total mass of the black hole at the final stage
of Penrose's process of energy extraction, considering that
the maximum possible energy is extracted. It is also related
to the energy contained within the external event horizon
$E(r_+)$ of the black hole (the surface of constant radius
$r=r_+$ defines the external event horizon). Every expression
for local or quasi-local gravitational energy must necessarily
yield the value of $E(r_+)$ in close agreement with $2M_{irr}$,
since we know beforehand the value of the latter as a function
of the initial angular momentum of the black hole\cite{Chris}.
The evaluation of $2M_{irr}$ is a
crucial test for any expression for the gravitational energy.
$E(r_+)$ has been obtained by means of different energy
expressions in Ref. \cite{Bergqvist}. Our expression for the
gravitational energy is the only one  that yields a
satisfactory value for $E(r_+)$, strikingly close to
$2M_{irr}$, and that arises in the framework of the
Hamiltonian formulation of the gravitational field.

In the Hamiltonian formulation of the TEGR\cite{Maluf3}
there arises a set of primary constraints $\Gamma^{ik}$
that satisty the angular momentum algebra. Following the
prescription for defining the gravitational energy,
the definition of the gravitational angular momentum arises
by suitably interpreting the integral form of the
constraint equation $\Gamma^{ik}=0$ as an angular momentum
equation. We apply this definition to the
gravitational field of a thin, slowly rotating mass shell.
In the limit of slow rotation we obtain a realistic measure
of the angular momentum of the field in terms of the moment
of inertia of the source.\par

\bigskip
\noindent Notation: space-time indices $\mu, \nu, ...$ and SO(3,1)
indices $a, b, ...$ run from 0 to 3. Time and space indices are
indicated according to
$\mu=0,i,\;\;a=(0),(i)$. The tetrad field $e^a\,_\mu$ 
yields the definition of the torsion tensor:  
$T^a\,_{\mu \nu}=\partial_\mu e^a\,_\nu-\partial_\nu e^a\,_\mu$.
The flat, Minkowski space-time  metric is fixed by
$\eta_{ab}=e_{a\mu} e_{b\nu}g^{\mu\nu}= (-+++)$.        \\

\bigskip
\bigskip

\noindent{\bf II. The Hamiltonian constraint equation as an
energy equation for the gravitational field}\par
\bigskip

We summarize here the Hamiltonian formulation obtained in
Ref. \cite{Maluf1}, where Schwinger's time gauge is assumed.
The Hamiltonian density constructed out of triads $e_{(i)j}$
restricted to the three-dimensional spacelike hypersurface, and
of the momenta canonically conjugated $\Pi^{(i)j}$, is given by

$$H=NC+N^iC_i+\Sigma_{mn}\Pi^{mn}+
{1\over {8\pi G}}\partial_k(NeT^k)+\partial_k(\Pi^{jk}N_j)
\;,\eqno(2.1)$$

\noindent where $N$ and $N^i$ are lapse and shift functions,
$\Sigma_{mn}=-\Sigma_{nm}$ are Lagrange multipliers,
$G$ is the gravitational constant and
$\Pi^{ij}=e_{(k)}\,^i \Pi^{(k)j}$.
The constraints are defined by

$$C=\partial_j(2keT^j)-ke\Sigma^{kij}T_{kij}
-{1\over{4ke}}\biggl(\Pi^{ij}\Pi_{ji}-{1\over 2}\Pi^2\biggr)\;,
\eqno(2.2)$$

$$C_k=-e_{(j)k}\partial_i\Pi^{(j)i}-\Pi^{(j)i}T_{(j)ik}
\;,\eqno(2.3)$$

\noindent where $e=det(e_{(i)j})$ and $k={1\over {16\pi G}}$. 
The tensor $\Sigma^{kij}$ reads

$$\Sigma^{kij}={1\over 4}(T^{kij}+T^{ikj}-T^{jki})+
{1\over 2}(g^{kj}T^i-g^{ki}T^j)\;.\eqno(2.4)$$

\noindent The trace of the torsion tensor is 
$T^i=g^{ik}T_k=g^{ik}e^{(m)j}T_{(m)jk}$.
The definition of $\Sigma^{kij}$ yields

$$\Sigma^{kij}T_{kij}={1\over 4}T^{kij}T_{kij}+
{1\over 2}T^{kij}T_{ikj}-T^iT_i\;.$$

The first two terms on the right hand side
of (2.2) are equivalent to the scalar curvature $R(e_{(i)j})$
on the three-dimensional spacelike hypersurface,

$$2\partial_j (eT^j)-e\Sigma^{kij}T_{kij}=
eR(e_{(i)j})\;.\eqno(2.5)$$

The integral form of the Hamiltonian constraint equation
$C(x)=0$ can be interpreted as an energy equation\cite{Maluf2},

$$\int d^3x\, \partial_j(2keT^j)=
\int d^3x \biggl\{ke\Sigma^{kij}T_{kij}+
{1\over{4ke}}\biggl(\Pi^{ij}\Pi_{ji}-{1\over 2}\Pi^2\biggr)
\biggr\}\;.\eqno(2.6)$$

\noindent We identify Eq. (2.6) as an energy equation because
the integral of the left
hand side of this equation over the whole three-dimensional
space yields the Arnowitt-Deser-Misner energy\cite{ADM},

$${1\over{8\pi G}}\int d^3 x \partial_j(eT^j)=
{1\over {16\pi G}}\int_S dS_k(\partial_i h_{ik}-
\partial_k h_{ii})=E_{ADM}\;.\eqno(2.7)$$

\noindent The right hand side of Eq. (2.7) is obtained by
requiring the asymptotic behaviour

$$e_{(i)j}\simeq \eta_{ij}+{1\over 2}h_{ij}({1\over r})
\;,\eqno(2.8)$$

\noindent in the limit $r \rightarrow \infty$. $\eta_{ij}$ is
the spatial sector of Minkowski's metric tensor and $h_{ij}$ is
the first term in the asymptotic expansion of $g_{ij}$.
Therefore we define the gravitational energy enclosed by a
volume $V$ of the three-dimensional space as\cite{Maluf2}

$$E_g={1\over{8\pi G}}\int_V d^3x\partial_j(eT^j)\;.\eqno(2.9)$$

\noindent The expression above has been applied to several
configurations of the gravitational field. The most relevant
application is the evaluation of the irreducible mass of the
Kerr black hole\cite{Maluf4}.\par
\bigskip
\bigskip

\noindent {\bf III. Gravitational energy expression in terms
of tetrad fields}\par
\bigskip
An expression for the gravitational energy density also arises
in the framework of the Hamiltonian formulation of general
relativity in the teleparallel geometry\cite{Maluf3}, without
posing any {\it a priori} restriction on the tetrad fields,
again interpreting the integral form of the constraint
equations as energy-momentum equations for the gravitational
field.

The Hamiltonian formulation developed in Ref. \cite{Maluf3}
is obtained from the Lagrangian density in empty space-time
defined by

$$L(e)\;=\;-k\,e\,\biggl( {1\over 4} T^{abc}T_{abc} +
{1\over 2}T^{abc}T_{bac}-T^aT_a\biggr)\;,\eqno(3.1)$$

\noindent where $e=det(e^a\,_\mu)$,
$T_{abc}=e_b\,^\mu e_c\,^\nu T_{a \mu \nu}$ and
the trace of the torsion tensor is given by 
$T_b=T^a\,_{ab}\;.$
The Hamiltonian is obtained by just rewriting the Lagrangian
density in the form $L=p\dot q -H$.
It has not been made use of any kind of projection of metric
variables to the three-dimensional spacelike hypersurface.
Since there is no time
derivative of $e_{a0}$ in (3.1), the corresponding momentum
canonically conjugated $\Pi^{a0}$ vanishes identically.
Dispensing with surface terms the total Hamiltonian density
reads\cite{Maluf3} 

$$H(e_{ai},\Pi^{ai})
=e_{a0}C^a+\alpha_{ik}\Gamma^{ik}+\beta_k\Gamma^k\;,\eqno(3.2)$$

\noindent where $\lbrace C^a, \Gamma^{ik}$ and $\Gamma^k\rbrace$
constitute a set of primary constraints, and
$\alpha_{ik}$ and $\beta_k$ are Lagrange  multipliers.
Explicit details are given in Ref. \cite{Maluf3}.
The first term of the constraint $C^a$ is given by a total
divergence in the form
$C^a=-\partial_k \Pi^{ak}+\cdot\cdot\cdot\; $.
In similarity with Eq. (2.6) we identify this total
divergence on the three-dimensional spacelike hypersurface
as the energy-momentum density of the gravitational field.
The total energy-momentum is defined by

$$P^a=-\int_V d^3x\,\partial_i \Pi^{ai}\;,\eqno(3.3)$$

\noindent where $V$ is an arbitrary space volume. It is invariant
under coordinate transformations on the spacelike manifold, and
transforms as a vector under the global SO(3,1) group
(we will return to this point later on). The
definition above generalizes expression (2.9) to tetrad fields
that are not restricted by the time gauge condition. However,
both expressions are equivalent, as we will see ahead,
if the time gauge condition is imposed.
After implementing the primary constraints $\Gamma^{ik}$ and
$\Gamma^k$, the expression of the momenta $\Pi^{ak}$ reads

$$\Pi^{ak}\;=\;k\,e\biggl\{ 
g^{00}(-g^{kj}T^a\,_{0j}-
e^{aj}T^k\,_{0j}+2e^{ak}T^j\,_{0j})$$

$$+g^{0k}(g^{0j}T^a\,_{0j}+e^{aj}T^0\,_{0j})
\,+e^{a0}(g^{0j}T^k\,_{0j}+g^{kj}T^0\,_{0j})
-2(e^{a0}g^{0k}T^j\,_{0j}+e^{ak}g^{0j}T^0\,_{0j})$$

$$-g^{0i}g^{kj}T^a\,_{ij}+e^{ai}(g^{0j}T^k\,_{ij}-
g^{kj}T^0\,_{ij})-2(g^{0i}e^{ak}-g^{ik}e^{a0})
T^j\,_{ji} \biggr\}\;.\eqno(3.4)$$

With appropriate boundary conditions expression (3.3) yields
the ADM energy. Let us consider asymptotically flat space-times
and assume that in the limit $r \rightarrow \infty$ the tetrad
fields have the asymptotic behaviour

$$e_{a\mu} \simeq \eta_{a\mu}
+ {1\over 2}h_{a\mu}({1\over r})\;,\eqno(3.5)$$

\noindent where $\eta_{a\mu}$ is Minkowski's metric tensor and
$h_{a\mu}$ is the first term in the asymptotic
expansion of $g_{\mu \nu}$.  Asymptotically flat space-times
are defined by Eq. (3.5) together with
$\partial_\mu g_{\lambda \nu}=O({1\over r^2})$,
or $\partial_\mu e_{a\nu}=O({1\over r^2})$. Considering
the $a=(0)$ component in Eq. (3.3) and integrating over
the whole three-dimensional spacelike hypersurface we find,
after a long but straightforward calculation, that

$$P^{(0)}=E = - \int_{V\rightarrow \infty} d^3x \partial_k
\Pi^{(0)k} =-2k\int_{V \rightarrow \infty} d^3x
\partial_k(eg^{ik}e^{(0)0}T^j\,_{ji})$$

$$={1\over {16\pi G}}\int_{S\rightarrow \infty}dS_k(\partial_i
h_{ik}-\partial_k h_{ii}) = E_{ADM}\;.\eqno(3.6)\;$$

We will prove that expressions (2.9) and (3.3)
coincide if we require the time gauge condition. In order to
prove it, let us rewrite $\Pi^{(0)k}$ as

$$\Pi^{(0)k}=e^{(0)}\,_i\Pi^{(ik)}
+e^{(0)}\,_i\Pi^{\lbrack ik \rbrack}
+e^{(0)}\,_0\Pi^{0k}\;,\eqno(3.7)$$

\noindent where  $(..)$ and $\lbrack .. \rbrack$ denote
symmetric and anti-symmetric components, respectively.
In the time gauge condition we have
$e_{(j)}\,^0=e^{(0)}\,_i=0$, and therefore
Eq. (3.7) reduces to $\Pi^{(0)k}=e^{(0)}\,_0\Pi^{0k}$.
An expression for $\Pi^{0k}$ can be obtained by requiring the
vanishing of the constraint $\Gamma^k$\cite{Maluf3},

$$ \Gamma^k \;=\;\Pi^{0k}\; +\;2k\,e\, (
g^{kj}g^{0i}T^0\,_{ij}-g^{0k}g^{0i}T^j\,_{ij}
+g^{00}g^{ik}T^j\,_{ij} )
\;.\eqno(3.8)$$

\noindent In the time gauge we have $T^0\,_{ij}=0$ and
therefore from $\Gamma^k=0$ we arrive at

$$\Pi^{0k}=
2ke(g^{0k}g^{0i}-g^{00}g^{ik})T^j\,_{ij}\;.\eqno(3.9)$$

\noindent All quantities in Eq. (3.9) are four-dimensional field
quantities. Let us now rewrite Eq. (3.9) in terms of field
quantities restricted to the three-dimensional spacelike
hypersurface by means of the lapse and shift functions, $N$ and
$N^i$, respectively. In view of the relations $e=N\,^3e$,$\;\;$
$g^{0i}=N^i / N^2$ and
$g^{ik}=\,^3g^{ik}-(N^iN^k) / N^2$, Eq. (3.9) can be written as

$$\Pi^{0k}={2\over N}k\;(\,^3e) (\,^3g^{ik}) T^j\,_{ij}\;.$$

\noindent The superscript 3 indicates that the quantity is
projected  on the spacelike hypersurface.
Note that $T^j\,_{ij}$ is still given in terms of
four-dimensional field quantities.

We make use of a 3+1 decomposition for the tetrad fields
according to $e^{ai}=\,^3e^{ai}+(N^i / N)\eta^a$,
$\;e^a\,_i=\,^3e^a\,_i$, $\;\eta^a=-Ne^{a0}$ and
$e^a\,_0=\eta^a N+\,^3e^a\,_i N^i$. The tetrad
fields $\,^3e_{ai}$ and $\,^3e^{ai}$ are related to each other
by means of the metric tensor $g_{ij}$ and its inverse
$\,^3g^{ij}$.
With the help of these relations we can rewrite $T^j\,_{ij}$
in terms of quantities on the spacelike hypersurface in the
time gauge condition, in which case we have
$\eta^a=\delta^a_{(0)}$ and $e^{(0)}\,_0=N$. We eventually
arrive at

$$\Pi^{(0)k}=2keg^{ik}g^{jm}e^{(l)}\,_m T_{(l)ij}\;,\eqno(3.10)$$

\noindent where we have eliminated the superscript 3. It is
straightfoward to verify that $\Pi^{(0)k}=-T^k$, where $T^k$ and
$T_{(l)ij}$ are precisely the same quantities that appear in
section II, and in particular in expression (2.9). Therefore in
the time gauge condition we have

$$P^{(0)}=-\int_V d^3x\,\partial_i \Pi^{(0)i}=
{1\over{8\pi G}}\int_V d^3x\partial_j(eT^j)\;.\eqno(3.11)$$

Differently from the quasi-local energy expressions\cite{Brown},
Eq. (3.11) is an integral of a scalar
density over finite space volumes, which can be
transformed into a surface integral.
Therefore our expression is not bound, in principle, to
belong to any class of quasi-local energies. There is
no need of subtraction terms in the present framework.
And yet Eq. (3.11) does satisfy the usual
requirements for a quasi-local energy expression. According
to the latter requirements the quasi-local energy expression must
(i) vanish for the Minkowski space-time; (ii) yield the ADM and
Bondi mass in the appropriate limits; (iii) yield the
appropriate value for weak and spherically symmetric gravitational
fields and (iv) yield the irreducible mass of the Kerr black hole.
The Bondi energy in the TEGR has been discussed in Ref.
\cite{Maluf5}, and the latter requirement is discussed in
section V.

\bigskip
\bigskip


\noindent {\bf IV. The determination of tetrad fields}\par
\bigskip

In the framework of the teleparallel geometry the gravitational
field can be described by an anholonomic transformation between
a reference space-time and the physical space-time. We will
briefly recall the difference between holonomic and
anholonomic transformations. Let us consider  two sets of
coordinates, $q^a=(t,x,y,z)$ and $x^\mu=(t,r,\theta,\phi)$,
related by the coordinate transformation
$dq^a=e^a\,_\mu dx^\mu$ such that

$$e^a\,_\mu={{\partial q^a}\over{\partial x^\mu}}=
\pmatrix{1&0&0&0\cr
0&\sin\theta\, \cos\phi & r\,\cos\theta\,\cos\phi
& -r\,\sin\theta\,\sin\phi\cr
0&\sin\theta\, \sin\phi & r\,\cos\theta\,\sin\phi
&  r\,\sin\theta\,\cos\phi\cr
0&\cos\theta & -r\,\sin\theta & 0\cr}\;. \eqno(4.1)$$

\noindent The relation $dq^a=e^a\,_\mu dx^\mu$ can
be integrated over the whole space-time, and therefore the
transformation $q^a \rightarrow x^\mu$ corresponds to a
single-valued global transformation. In this case the
transformation is called holonomic and both coordinate sets
describe the same space-time.

However, in the general case the relation
$dq^a=e^a\,_\mu dx^\mu$ cannot be globally integrated, since
$e^a\,_\mu$ may not be a gradient function of the type
$\partial_\mu  q^a $. If the quantities
$e^a\,_\mu$ are such that
$\partial_\mu e^a\,_\nu - \partial_\nu e^a\,_\mu \ne 0$, then
the transformation is called anholonomic.

For the tetrads given by (4.1) the torsion tensor
$T^a\,_{\mu\nu}=\partial_\mu e^a\,_\nu - \partial_\nu e^a\,_\mu$
vanishes. It is known that $T^a\,_{\mu\nu}$ vanish
identically if and only if $e^a\,_\mu$ are gradient
vectors\cite{Schouten2}.
In the framework of the TEGR the gravitational field corresponds
to a configuration such that $T^a\,_{\mu\nu}\ne 0.$ Thus every
gravitational field is  described by a space-time that is
anholonomically related to the four-dimensional Minkowski
space-time, which is taken as the reference space-time.
Consequently the tetrad fields to be considered must necessarily
yield a vanishing torsion tensor 
in the limit of vanishing physical parameters (such as mass,
angular momentum and charge), in which case the tetrad field
must reduce to expression (4.1), or, in the case of arbitrary
coordinates, to the form $e^a\,_\mu = \partial_\mu q^a$.

The idea of describing the gravitational field as the gauge
field of the Poincar\'e group is rather widespread. In view of
the general acceptance of this idea, there is a unjustified
prejudice against gravitational theories that do not exhibit
local SO(3,1) symmetry. Rather than being a drawback of the
present formulation, the requirement of a global set of tetrad
fields for the description of the space-time is a natural
feature of teleparallel theories\cite{Einstein} and of the
teleparallel geometry.

Before addressing the problem of obtaining the appropriate set
of tetrad fields out of a given metric tensor, it is
instructive to analyze the construction of tetrads for the flat
space-time, since a number of features that take place in this
context carry over to the general case of an arbitrary
space-time metric tensor. We will consider two sets of tetrads
that describe the flat space-time, and that reveal the
relationship between the reference space-time with
coordinates $q^a$ and the physical space-time with coordinates
$x^\mu$.

For a given arbitrary function $\omega(t)$ let us consider a
transformation between two rotating cartesian coordinate
systems, $q^0=t$, $q^1=x^1\cos\omega(t)-x^2\sin\omega(t)$,
$q^2=x^1\sin\omega(t)+x^2\cos\omega(t)$, $q^3=x^3$. The tetrads
are given by

$$e^a\,_\mu(t,x,y,z)=
\pmatrix{1&0&0&0\cr
-(x^1\sin\omega+x^2\cos\omega)\dot\omega
&\cos\omega&-\sin\omega &0\cr
(x^1\cos\omega-x^2\sin\omega)\dot\omega
&\sin\omega &\cos\omega&0\cr
0&0&0&1}\;.\eqno(4.2)$$

\noindent These tetrads describe a flat space-time with
cartesian coordinates $x^\mu$ that is rotating with respect to
the reference space-time with coordinates $q^a$. We notice the
appearance of anti-symmetric components in the spatial sector
of $e^a\,_\mu$. This is a general feature in cartesian
coordinates: under an infinitesimal rotation a rotated vector
$\tilde V$ is related to the vector $V$ by means of the relation
$\tilde V=RV$; the rotation matrix is given by
$R=1+\omega_iX^i$, where $\omega_i$ are arbitrary parameters
and the generators $X^i$ are anti-symmetric matrices. Therefore
the emergence of anti-symmetric components in the sector
$e_{(i)j}(t,x,y,z)$ is expected if the two space-times are
rotating with respect to each other.

Another transformation of general character is a Lorentz boost,
$q^{(0)}=\gamma(t+(v/c^2)x^1)$, $q^{(1)}=\gamma(x^1+vt)$,
$q^{(2)}=x^2$ and $q^{(3)}=x^3$, where
$\gamma=1/\sqrt{1-v^2/c^2}$ (assuming the velocity of light
$c\ne 1$). The two space-times have different time scales.
The tetrads read

$$e^a\,_\mu(t,x,y,z)=
\pmatrix{ \gamma&(v/c^2)\gamma &0&0\cr
v\gamma &\gamma&0&0\cr
0&0 &1&0\cr
0&0&0&1}\;.\eqno(4.3)$$

\noindent The tetrads above do not satisfy the time gauge
condition because of the emergence of the term
$e^{(0)}\,_1=(v/c^2)\gamma$. Under an arbitrary boost
transformation there will arise terms such that
$e^{(0)}\,_k\ne 0$, which violate the time gauge condition
$e_{(i)}\,^0=0$. The main feature of the time gauge condition
is to lock the time axes of the reference space-time and of
the physical space-time.

In the absence of the gravitational field,
$e^a\,_\mu(t,x,y,z)=\delta^a_\mu$ is the unique set of tetrads
that describes a reference space-time with coordinates $q^a$
that is neither related by a boost transformation nor rotating
with respect to the physical space-time with coordinates
$x^\mu$. The features above should also carry over to the case
of an arbitrary gravitational field. As we will see, they are
essential in the description of the energy properties of the
gravitational field. Likewise, for a given space-time metric
tensor the set of tetrad fields that in cartesian
coordinates satisfy the properties

$$e_{(i)j}=e_{(j)i}\;,\eqno(4.4a)$$

$$e_{(i)}\,^0=0\,,\eqno(4.4b)$$

\noindent establish a {\it unique} reference space-time that
is neither related by a boost transformation, nor rotating
with respect to the physical space-time. Equations
(4.4b) fix six degrees of freedom of the tetrad field.
The importance of Eqs.
(4.4a,b) to the definition of the gravitational energy will be
discussed at the end of section V.

Let us consider now the Kerr space-time.
In terms of Boyer-Lindquist\cite{BL} coordinates the Kerr metric
tensor is given by

$$ds^2=-{\psi^2 \over\rho^2}dt^2-
{{2\chi \sin^2\theta}\over \rho^2}d\phi\,dt
+{\rho^2 \over \Delta}dr^2+
\rho^2 d\theta^2+
{{\Sigma^2 \sin^2\theta}\over \rho^2}d\phi^2 \;,\eqno(4.5)$$

\noindent where $\rho^2=r^2+a^2 cos^2\theta$, 
$\Delta= r^2 +a^2 -2mr$, $\chi=2amr$ and

$$\Sigma^2=(r^2+a^2)^2-\Delta a^2\,\sin^2\theta\;,$$

$$\psi^2=\Delta -a^2\,\sin^2\theta\;.$$

Each set of tetrad fields defines a teleparallel geometry.
For a given space-time metric tensor $g_{\mu\nu}$, there exists
an infinite set of tetrad fields that yield $g_{\mu\nu}$. From
the point of view of the metrical properties of the space-time,
any two set of tetrads out of this infinity corresponds to
viable (but distinct) teleparallel configurations\cite{Nester2}.
However, the description of the gravitational field energy
requires at least boundary conditions. In the framework of the
teleparallel geometry the correct description of the
gravitational energy-momentum singles out a unique set of tetrad
fields. In the following we will consider the most relevant
tetrad configurations. The first one is based on the weak field
approximation first suggested by M\o ller, given by expression
(3.5),

$$e^M_{a\mu} \simeq \eta_{a\mu}
+ {1\over 2}h_{a\mu}\;,\eqno(4.6a)$$

\noindent together with the symmetry condition on $h_{a\mu}$,

$$h_{a\mu}=h_{\mu a}\;.\eqno(4.6b)$$

\noindent Note that Eq. (4.6a) is demanded not only in the
asymptotic limit, but at every space-time point. Although the
weak field limit fixes the expression of $e^M_{a\mu}$,
the resulting expression is taken to hold in the strong field
regime. The expression that satisfies Eq. (4.6) and that
yields Eq. (4.5) is given by

$$e^M_{a\mu}=\pmatrix{
-{\psi \over \rho}\sqrt{1+M^2y^2} & 0 & 0 &
-{{\chi N y}\over {\psi \rho}}\sin^2\theta\cr
{{\chi y}\over {\Sigma \rho}}\sin\theta\,\sin\phi &
{\rho \over \sqrt{\Delta}}\sin\theta\,\cos\phi &
\rho\, \cos\theta\,\cos\phi &
-{\Sigma \over \rho}\sqrt{1+M^2N^2y^2}\sin\theta\,\sin\phi\cr
-{{\chi y}\over{\Sigma \rho}}\sin\theta\, \cos\phi &
{\rho \over \sqrt{\Delta}} \sin\theta\,\sin\phi &
\rho\, \cos\theta\,\sin\phi &
{\Sigma \over \rho}\sqrt{1+M^2N^2y^2}\sin\theta\,\cos\phi \cr
0 & {\rho \over \sqrt{\Delta}}\cos\theta &
-\rho\, \sin\theta & 0 \cr}\,,\eqno(4.7) $$

\noindent where

$$y^2={{ 2N\sqrt{1+M^2}-(1+N^2)}\over
{4M^2N^2-(1-N^2)^2}}\;,$$

$$M={\chi \over{\Sigma \psi}}\sin\theta\;,$$

$$N={{\psi r}\over \Sigma}\;.$$

The second set of tetrad fields to be considered satisfies
the weak field approximation

$$e_{(i)j}\simeq \eta_{ij}+{1\over 2} h_{ij}\;,\eqno(4.8a)$$

$$h_{ij}=h_{ji}\;,\eqno(4.8b)$$

\noindent together with Schwinger's time gauge condition,
$e_{(k)}\,^0=e^{(0)}\,_j=0$ (Eq. (4.4b)). Note that Eqs.
(4.8a,b) are essentially equivalent to Eq. (4.4a). Conditions
(4.8) are assumed to fix the expression of
$e^a\, _\mu$ also in the strong  field regime. The set of
tetrad fields that satisfies Eqs. (4.8), (4.4b) and
that yields Eq. (4.5) reads

$$e^S_{a\mu}=\pmatrix{
-{1\over \rho}\sqrt{\psi^2+{\chi^2\over \Sigma^2}\sin^2\theta} &
0&0&0\cr
{\chi \over {\Sigma \rho}}\sin\theta\,\sin\phi &
{\rho \over \sqrt{\Delta}}\sin\theta\,\cos\phi &
\rho\,\cos\theta\,\cos\phi &
-{\Sigma \over \rho} \sin\theta\,\sin\phi\cr
-{\chi \over{\Sigma \rho}}\sin\theta\,\cos\phi &
{\rho \over \sqrt{\Delta}}\sin\theta\,\sin\phi &
\rho\,\cos\theta\,\sin\phi &
{\Sigma \over \rho}\sin\theta\,\cos\phi\cr
0&{\rho \over \sqrt{\Delta}}\cos\theta&-\rho\,\sin\theta&0\cr}
\,.\eqno(4.9)$$

\noindent We note finally that both Eqs. (4.7) and (4.9)
reduce to Eq. (4.1) if we make $m=a=0$.\par
\bigskip
\bigskip

\noindent {\bf V. The irreducible mass of the Kerr black
hole}\par
\bigskip

In this section we will apply expression (3.3) to the evaluation
of the irreducible mass $M_{irr}$ of the Kerr black hole. This
is the most important test for any gravitational energy
expression, local or quasi-local, since the geometrical setting
corresponds to an intricate configuration of the gravitational
field, and since the value of $M_{irr}$ is known from the
work of Christodoulou\cite{Chris}.

In order to obtain $M_{irr}$ we will calculate the $a=(0)$
component of Eq. (3.3) by fixing $V$ to be the volume within
the $r= r_+$ surface, where $r_+=m+\sqrt{m^2-a^2}$ is the
external horizon of the Kerr black hole. Therefore we will
consider

$$P^{(0)}=E=-\int_S dS_i \,\Pi^{(0)i}=
-\int_S d\theta d\phi \,\Pi^{(0)1}(r,\theta,\phi) \;,\eqno(5.1)$$

\noindent where the surface $S$ is determined by the condition
$r= r_+$. The expression of $\Pi^{(0)1}$ will be obtained
by considering Eq. (4.7). In view of Eq. (3.11) there is no
need to calculate Eq. (5.1) out of Eq. (4.9), since it has
already been evaluated\cite{Maluf4}.

In the Appendix we present the expressions of the components of
the torsion tensor constructed out of the tetrad configuration
Eq. (4.7). The component $\Pi^{(0)1}$ is then obtained from the
definition (3.4) by means of simple (albeit long) algebraic
manipulations. The expression of $\Pi^{(0)1}(r,\theta,\phi)$
for the tetrad expression (4.7) reads

$$\Pi^{(0)1}={{k\Sigma y}\over \rho}\sin\theta\,\biggl\{
-2\biggl(1+N\Omega +{\rho^2 \over{y\Sigma}}\biggr)
+{{2\sqrt{\Delta}}\over \Sigma}\partial_r \Sigma
+{{\sqrt{\Delta} N}\over \Omega}\biggl(
{M^2\over \chi}\partial_r \chi
+ {2\over \Sigma}\partial_r \Sigma\biggr) \biggr\}\;,\eqno(5.2)$$

\noindent where the definitions of $y, N$ and $M$ are given
after expression (4.7) and

$$\Omega=\sqrt{1+M^2}\;.$$

\noindent  On the surface $r=r_+$ we have
$\Delta(r_+)=0$, $M^2(r_+)=-1$ and $\Omega(r_+)=0$. Therefore
the last term in Eq. (5.2) is indefinite. It must be calculated
by taking the limit $r\rightarrow r_+$. We find

$$\lim_{r\rightarrow r_+}
{{\sqrt{\Delta} N}\over \Omega}\biggl(
{M^2\over \chi}\partial_r \chi
+ {2\over \Sigma}\partial_r \Sigma\biggr)=
-{{a^2\sin^2\theta}\over m}\biggl(
{\sqrt{m^2-a^2}\over {2mr_+}}+
{r_+ \over{2mr_+ -a^2\sin^2\theta}}\biggr)\;.$$

\noindent The other terms in Eq. (5.2) do not pose any problem,
and thus we can obtain the expression of the energy
contained within the external event horizon of the Kerr black
hole, that follows from the tetrad configuration (4.7). The final
expression arises as a function of the angular momentum per unit
mass $a$. It is given by (we are assuming $G=1$)

$$E\lbrack e^M_{a\mu}\rbrack = {m\over 4}
\int_0^\pi d\theta\,\sin\theta\,\bigg[
\sqrt{p^2+\lambda^2 \cos^2\theta}+
{{py}\over \sqrt{p^2+\lambda^2 \cos^2\theta}}$$

$$+{ {2p^3y} \over{(p^2+\lambda^2 \cos^2\theta)^{3\over 2}}}-
{ {y(p-1) \sqrt{p^2+\lambda^2 \cos^2\theta}}\over 2}
\biggr]\;,\eqno(5.3)$$

\noindent where

$$p=1+\sqrt{1-\lambda^2}\;\;\;\;,\;\;\;\;
a=\lambda m,\;\;\;\; 0\leq \lambda \leq 1\;.$$

For the tetrad configuration Eq. (4.9) we have\cite{Maluf4}

$$E\lbrack e^S_{a\mu}\rbrack =
m\biggl[ {\sqrt{2p}\over 4}+{{6p-\lambda^2}\over 4\lambda} \ln
\biggl( {{\sqrt{2p} +\lambda}\over p} \biggr) \biggr]
\;.\eqno(5.4)$$

Expressions (5.3) and (5.4) must be compared with $2M_{irr}$,
where $M_{irr}$ is given by\cite{Chris}
$M_{irr}={1\over 2}\sqrt{{r_+}^2+a^2}$. In our notation we have

$$2M_{irr}=m\sqrt{2p}\;.\eqno(5.5)$$

\noindent In the limit $a\rightarrow 0$ all energy expressions
yield $2m$, which is the value obtained by several different
approaches\cite{Bergqvist}. In
figure 1 we have plotted $\varepsilon =E / m$ against $\lambda$,
where $0\le \lambda \le 1$. Each value of $\lambda$
characterizes an angular momentum state of the black hole.
The hope was that the tetrad field given by Eq. (4.7) would
explain the tiny difference between the numerical
values of Eqs. (5.4) and (5.5). However,
the deviation of expression (5.3) from $2M_{irr}$ indicates that
the tetrad configuration Eq. (4.7) is not appropriate to the
description of gravitational energy. The latter is most correctly
described by requiring the tetrad configuration Eq. (4.9), that
satisfies Schwinger's time gauge condition together with Eq.
(4.4a).

The choice of the tetrad field given by Eq. (4.9) amounts to
choosing the unique reference space-time that is neither related
by a boost transformation nor rotating with
respect to the physical space-time.

For an arbitrary space volume $V$ the gravitational energy is
defined relationally, in the sense that it depends on the choice
of the reference space-time. If the tetrad fields are required
to satisfy conditions (4.4a,b) for a metric tensor that
exhibits asymptotic boundary conditions similar to Eq. (3.5),
then for asymptotically flat space-times the physical space-time
coincides with the reference space-time in the limit
$r \rightarrow \infty$.
If, however, we choose a reference space-time that is, for
instance, rotating (about the $z$ axis, say) with respect to the
space-time defined by the Kerr solution, then the
irreducible mass of the black hole, calculated with respect to
this reference space-time, will be different from expression
(5.4), the difference residing in rotational effects. Therefore
in similarity to the ordinary concept of energy, the
gravitational energy depends on the rotational state of the
reference frame. Rotational and boost effects are eliminted by
requiring conditions (4.4a,b) on the tetrad fields.

The agreement between Eqs. (5.4) and (5.5) is the most important
result so far obtained from definitions (2.9) and (3.3). To our
knowledge, the latter are the only energy definitions that
yield a value satisfactorily close to $2M_{irr}$, and that arise
from the structure of the Hamiltonian formulation of the
theory.

Before closing this section we note that the time gauge
condition (4.4b) breaks the SO(3,1) symmetry group into
the global SO(3). Therefore in this case $P^a$ given by
Eq. (3.3) is no longer a true SO(3,1) vector. \par


\vskip 2.0cm
\noindent {\bf VI. Angular momentum of the gravitational
field}\par
\bigskip

In the context of Einstein's general relativity rotational
phenomena is certainly not a completely understood issue. The
prominent manifestation of a purely relativistic rotational
effect is the dragging of inertial frames. If the angular
momentum of the gravitational field of isolated systems has a
meaningful notion, then it is reasonable to expect the latter
to be somehow related to the rotational motion of the physical
sources.

The angular momentum of the gravitational field has been
addressed in the literature by means of different
approaches. The oldest approach is based on 
pseudotensors\cite{LL,BT}, out of which angular momentum
superpotentials are constructed. An alternative approach assumes
the existence of certain Killing vector fields that allow the
construction of conserved integral quantities\cite{Komar}.
Finally, the gravitational angular momentum can also be
considered in the context of Poincar\'e gauge theories of
gravity\cite{PGT}, either in the Lagrangian or in the Hamiltonian
formulation. In the latter case it is required that the
generators of spatial rotations at infinity have well defined
functional derivatives. From this requirement a certain surface
integral arises, whose value is interpreted as the gravitational
angular momentum. 

The main motivation for considering the angular momentum of the
gravitational field in the present investigation resides in the
fact that the constraints $\Gamma^{ik}$\cite{Maluf3},

$$\Gamma^{ik}=-\Gamma^{ki}=
2\Pi^{[ik]}-2\,k\,e \biggl( -g^{im}g^{kj}T^0\,_{mj}+
(g^{im}g^{0k}-g^{km}g^{0i})T^j\,_{mj} \biggr)
\;,\eqno(6.1)$$

\noindent satisfy the angular momentum algebra,

$$\lbrace \Gamma^{ij}(x),\Gamma^{kl}(y)\rbrace=\biggl(
g^{il}\Gamma^{jk}+g^{jk}\Gamma^{il}-
g^{ik}\Gamma^{jl}-g^{jl}\Gamma^{ik}\biggr)\delta(x-y)
\;,\eqno(6.2)$$

Following the prescription for defining the gravitational energy
out of the Hamiltonian constraint of the TEGR, we interpret the
integral form of the constraint equation $\Gamma^{ik}=0$ as an
angular momentum equation, and therefore we define the angular
momentum of the gravitational field $M^{ik}$ according to

$$M^{ik}=2\int_V d^3x\, \Pi^{[ik]} 
= 2k\int_V d^3x\, e \biggl[ -g^{im}g^{kj}T^0\,_{mj}+
(g^{im}g^{0k}-g^{km}g^{0i})T^j\,_{mj} \biggr]
\;,\eqno(6.3)$$

\noindent for an arbitrary volume $V$ of the three-dimensional
space.

In Einstein-Cartan type theories there also appear constraints
that satisfy the Poisson bracket given by Eq. (6.2). However, such
constraints arise in the form $\Pi^{[ik]}=0$, and so a definition
similar to Eq. (6.3), i.e., interpreting the constraint equation
as an equation for the angular momentum of the field, is not
possible.

Since definition (6.3) is a
three-dimensional integral we will consider a
non-singular space-time metric that exhibits rotational motion.
One exact solution that is everywhere regular in the exterior
and interior regions of the rotating source is the 
metric associated to a thin, slowly rotating mass shell as
described by Cohen\cite{Cohen}. In the limit of small angular
momentum this metric corresponds to the asymptotic form of
Kerr's metric tensor. The main motivation for
considering this metric is the construction of a realistic
source for the exterior region of the Kerr space-time, and
therefore to match the latter region to a singularity-free
space-time. For a shell of radius $r_0$ and total mass
$m=2\alpha$ as seen by an observer at infinity, the metric
reads

$$ds^2=-V^2dt^2+\psi^4\lbrack dr^2+r^2d\theta^2+
r^2\sin^2\theta(d\phi-\Omega dt)^2\rbrack\;,\eqno(6.4)$$

\noindent where

$$V={{ r_0-\alpha}\over{r_0 + \alpha}}\;,$$

$$\psi = \psi_0 =1+ {\alpha \over r_0}\;\;\;\;,\;\;\;\;
\Omega=\Omega_0=const.\;,$$

\noindent for $r < r_0$, and

$$V={{r-\alpha}\over{r+\alpha}}\;,$$

$$\psi=1+{\alpha \over r}\;\;\;\;,\;\;\;\;
\Omega=\biggl({{r_0 \psi_0^2}\over{r \psi^2}}\biggr)^3
\Omega_0\;,$$

\noindent for $r> r_0$.

The set of tetrad fields that satisfy conditions (4.4a,b) is
given by

$$e_{a\mu}=\pmatrix{ -V&0&0&0\cr
\Omega r\psi^2 \sin\theta\,\sin\phi &
\psi^2 \,\sin\theta\,\cos\phi &
r\psi^2\,\cos\theta\,\cos\phi &
-r\psi^2\,\sin\theta\,\sin\phi\cr
-\Omega r\psi^2 \sin\theta\,\cos\phi &
\psi^2 \,\sin\theta\,\sin\phi &
r\psi^2\,\cos\theta\,\sin\phi &
r\psi^2\,\sin\theta\,\cos\phi\cr
0 & \psi^2\,\cos\theta & -r\psi^2\,\sin\theta & 0 \cr}
\;.\eqno(6.5)$$

\noindent The determinant of $e_{a\mu}$ is
$e=V r^2\psi^6\,\sin\theta$.

The nonvanishing components of the torsion tensor that are
needed in the following read

$$T^{(1)}\,_{12}= r\partial_r \psi^2\,\cos\theta\,\cos\phi\;,$$

$$T^{(2)}\,_{12}= r\partial_r \psi^2\,\cos\theta\,\sin\phi\;,$$

$$T^{(3)}\,_{12}= -r\partial_r \psi^2\,\sin\theta\;,$$

$$T^{(1)}\,_{13}= -r\partial_r \psi^2\,\sin\theta\,\sin\phi\;,$$

$$T^{(2)}\,_{13}= r\partial_r \psi^2\,\sin\theta\,\cos\phi\;.$$

\noindent The anti-symmetric components
$\Pi^{\lbrack ik \rbrack}$ can be easily evaluated. We obtain

$$\Pi^{\lbrack 13 \rbrack}(r,\theta,\phi)=
4k\alpha{\Omega \over V}\psi\,\sin\theta\;,$$

\noindent for $r>r_0$, 
$\Pi^{\lbrack 13 \rbrack}(r,\theta,\phi)=0$ for $r<r_0$, and
$\Pi^{\lbrack 12 \rbrack}(r,\theta,\phi)=
\Pi^{\lbrack 23 \rbrack}(r,\theta,\phi)=0$ for any value of
$r$. In cartesian coordinates the only nonvanishing component
of the total angular momentum is given by

$$M^{12}=2\int d^3x\,\Pi^{\lbrack 12\rbrack}(x,y,z)=
4\pi\int^\pi_0 d\theta \int_0^\infty dr\,r \,\sin^2\theta\,
\Pi^{\lbrack 13 \rbrack}
(r,\theta,\phi)$$

$$=\alpha \, \int_0^\pi d\theta\,\sin^3\theta 
\int_{r_0}^\infty dr\;r\psi {\Omega \over V}\;.\eqno(6.6)$$

The integral above is finite, well behaved and can be
exactly computed. However, we are interested only in the
limit $r_0 >>\alpha$, in which case Cohen identifies
$J=1/2 (r_0\psi_0^2)^3\Omega_0$ as the
Newtonian value for the  angular momentum of a rotating
mass shell\cite{Cohen}. In this limit the calculation is
straightforward. We find

$$M^{12} \simeq {{8\alpha}\over {3r_0}}J
={{4m}\over{3r_0}}J\;.\eqno(6.7)$$

We identify $M^{12}$ as the angular momentum of the
gravitational field. Substituting the expression of $J$ in
Eq. (6.7) and considering that in the limit $r_0>>\alpha$ we
have $\psi_0=1+\alpha/r_0 \simeq 1$, we arrive at

$$M^{12}=\biggl( {2\over 3} mr_0^2\biggr) \Omega_0\;.
\eqno(6.8)$$

\noindent $\Omega_0 =\Omega(r_0)$ is the induced angular
velocity of inertial frames inside the shell\cite{Brill}.
The term between the parentheses in the expression
above corresponds to the moment of inertia of a rotating
shell of radius $r_0$ and mass $m$. For
small $\alpha$, $\Omega_0$ and the angular velocity of the
shell $\omega_s$ are related via
$\Omega_0=\omega_s(4m/3r_0)$\cite{Brill}.
Therefore in the Newtonian limit $r_0>>\alpha$ we have
$M^{12}=({\Omega_0/\omega_s})J$, where
$J=(2/3)mr_0^2 \omega_s$.

The metric tensor (6.4) is likely to be the only exact solution
of Einstein's equations whose expression for the classical
angular momentum of the source is precisely known.

In order to assess the significance of the above result, we will
evaluate the angular momentum associated to the metric tensor
(6.4) by means of Komar's integral $Q_K$\cite{Komar},

$$Q_K={1\over {8\pi}}\oint_S
\sqrt{-g}\,\varepsilon_{\alpha \beta\mu\nu}
\nabla^{\lbrack \alpha} \xi^{\beta\rbrack}dx^\mu \wedge dx^\nu
\;,\eqno(6.9)$$

\noindent where $S$ is a spherical surface of radius
$R\rightarrow \infty$, $\xi^\mu$ is the Killing vector field
$\xi^\mu=\delta^\mu_3$ and $\nabla$ is the covariant derivative
constructed out of the Christoffel symbols
$\Gamma^\lambda_{\mu\nu}$. The integral $Q_K$ reduces to

$$Q_K={1\over {2\pi}}\oint_S\,\sqrt{-g}\,g^{0\mu}
\Gamma^1_{\mu 3}\,d\theta \,d\phi\;.\eqno(6.10)$$

\noindent By substituting Eq. (6.4) and taking the limit
$S\rightarrow \infty$ we obtain

$$Q_K={4\over 3}(r_0\psi_0^2)^3\Omega_0
\simeq {4\over 3} r_0^3\Omega_0={{16}\over 9}mr_0^2\omega_s
={8\over 3}J\;.\eqno(6.11)$$

\noindent In the equation above we are considering
$r_0>>\alpha$. We observe that definitions (6.3) and (6.9)
yield distinct results. In order to make clear the distinction
it is useful to rewrite both espressions, (6.7) and (6.11), in
laboratory (CGS) units. Thus we make $m=(G/c^2)M$ and
$\omega_s=\Omega_s/c$, where $M$ is given in grams, and
$\Omega_s$ in radians per second. In addition, we make the
replacement $1/(16\pi) \rightarrow c^3/(16\pi G)$ in the
multiplicative factor of both expressions, in order to yield
the correct dimension to the integrals.  We arrive at

$$M^{12}=\biggl({G\over c^2}\biggr){{4M}\over{3r_0}} \biggl(
{2\over 3} M r_0^2\Omega_s\biggr)\;,\eqno(6.7')$$

$$Q_K={8\over 3}\biggl({2\over 3}Mr_0^2\Omega_s\biggr)
\;.\eqno(6.11')$$

\noindent We note that $G/c^2=0,74\times 10^{-28} g/cm$. Both
expressions have angular momentum units.

One expects the gravitational angular momentum to be of the
order of magnitude of the intensity of the gravitational field.
We observe that Komar's integral yields a value proportional
to the angular momentum of the {\it source}, whereas $M^{12}$
is much smaller than $Q_K$. Indeed, the gravitational field of
a mass shell of typical laboratory values is negligible, and
consequently the gravitational angular momentum should be
negligible as well.  Therefore $M^{12}$ yields a realistic
value for the angular momentum of the gravitational field, in
contrast to $Q_K$.

The advantage of definition (6.4) is that it
does not depend on the existence of Killing vector fields.
The conclusion is that the angular momentum of the space-time
of a rotating mass shell, according to the definition
(6.3), is proportional to the induced angular velocity
$\Omega_0$ of inertial frames.

The investigations carried out so far in the context of the Kerr
solution are not yet conclusive. Although the
calculations in the Boyer-Lindquist coordinates are
extremely intricate, the indications are that $M^{12}$ diverges.
Considering the metric tensor given by Eq. (4.5) and the related
definitions, we calculate the anti-symmetric components of the
momenta $\Pi^{\lbrack ik \rbrack}$ in the time gauge, i.e., out
of tetrads (4.9). They are given by

$$\Pi^{\lbrack 12 \rbrack}(r,\theta,\phi)=0\;,\eqno(6.12a)$$

$$\Pi^{\lbrack 13 \rbrack}(r,\theta,\phi)=
{{k\chi\, \sin\theta}\over
\sqrt{\psi^2\Sigma^2+\chi^2\, \sin^2\theta}}
\biggl(1+{\rho^2\over \Sigma}-{\sqrt{\Delta}\over\Sigma}
\partial_r \Sigma\biggr)\;,\eqno(6.12b)$$

$$\Pi^{\lbrack 23 \rbrack}(r,\theta,\phi)={{k\chi}\over\sqrt{
\Delta (\psi^2\Sigma^2+\chi^2\, \sin^2\theta)}}
\biggl( \cos\theta\biggl({\rho^2\over \Sigma}-1\biggr)-
{{\sin\theta}\over \Sigma}\partial_\theta \Sigma\biggr)
\;.\eqno(6.12c)$$

\noindent Transforming to cartesian coordinates we obtain

$$M^{12}=\int d^3x \Pi^{\lbrack 12 \rbrack}(x,y,z)$$

$$=2\pi\int_0^\infty dr \int_0^\pi d\theta \biggl(
r\,\sin\theta\, \Pi^{\lbrack 13 \rbrack}(r,\theta,\phi)+
r^2\,\sin\theta\,\cos\theta\,
\Pi^{\lbrack 23 \rbrack}(r,\theta,\phi)
\biggr)\;,\eqno(6.13)$$

\noindent and $M^{13}=M^{23}=0$. The evaluation of Eq. (6.13)
out of expressions (6.12) yields a divergent result. The
latter is positively and negatively divergent in the external
($r_+$) and internal ($r_-$) horizons of the black hole,
respectively. Moreover, in the region $r_-<r<r_+$, $M^{12}$
acquires an imaginary component. A possible
interpretation is that the Boyer-Lindquist coordinates are not
suitable to the present analysis. In any way, integration over
the whole spacelike section of the Kerr space-time is a
nontrivial operation.

It must be noted that the Kerr black hole has no classical
analog. The interpretation of the angular momentum parameter
$a$ of the Kerr solution is not straightforward, since in 
the Newtonian theory of gravitation the gravitational field of a
body does not depend on its rotational motion. The parameter $a$
is identified with the angular momentum per unit mass of the
source only after reducing the exterior region of the Kerr
metric to a Lense-Thirring type metric by successive
approximations\cite{Adler}.

\bigskip
\bigskip
\noindent {\bf VII. Discussion}\par
\bigskip
In this paper we have investigated the definitions of energy
and angular momentum of the gravitational field
that arise in the Hamiltonian formulation of the TEGR.
We have  compared the most important achievement, i.e., the
calculation of the irreducible mass of the Kerr black hole,
with the result previously obtained in the framework of the
same theory, but with the Hamiltonian formulation established
under the {\it a priori} imposition of the
time gauge condition. The two results agreed. In fact, both
energy expressions coincide by requiring the time gauge
condition, if the latter is imposed {\it a posteriori} in
the $a=(0)$ component of expression (3.3).

The relevance of Eq. (5.4) is
further enhanced if we observe that the Brown-York
method\cite{Brown} for the evaluation of quasi-local
gravitational energy fails in obtaining a value close to the
irreducible mass of the Kerr black hole. Although the
calculations in the framework of this method are quite
intricate, recently it has been carried out\cite{DM}. It has
been shown that the gravitational energy within $r_+$ is close
to $2M_{irr}$ only for $a/m <0.5$ (fig. 1 of Ref. \cite{DM}).

Definitions for the gravitational energy in the context of
the teleparallel equivalent of general relativity have already
been proposed in the literature. In Ref. \cite{Nester1} an
expression for the gravitational energy arises from the surface
term of the total Hamiltonian (Eqs. (3.18) and (3.19) of Ref.
\cite{Nester1}). A similar quantity is suggested in Ref.
\cite{Blagojevic}, according to Eq. (3.8) of the latter
reference. Both expressions are equivalent to the integral
form of the total divergence of the Hamiltonian density developed
in Ref. \cite{Maluf3} (Eq. (27) of the latter reference),

$$E=\int_{V\rightarrow \infty}d^3x\, \partial_k
(e_{a0}\Pi^{ak})=\oint_{S\rightarrow \infty}
dS_k\,(e_{a0}\Pi^{ak}).$$

\noindent The three expressions yield the same value for the
{\it total} energy of the gravitational field. However, since
these three expressions contain the lapse function in the
integrand, none of them is suitable to the calculation of the
irreducible mass of the Kerr black hole, in which case we
consider a finite surface of integration,
because the lapse funtion vanishes on the external event
horizon of the black hole (recalling the 3+1 decomposition in
section III, $e^a\,_0=\eta^a N+\,^3e^a\,_iN^i $. In the time
gauge we have $\eta^a=\delta^a_{(0)}$ and $e^{(0)}\,_i=0$).
The energy expressions of Refs. \cite{Nester1,Blagojevic} are
not to be applied to a finite surface of integration;
rather, they yield the total energy of the space-time.

The energy expression (3.3) is defined with respect to a
given reference space. Tetrad fields that satisfy conditions
(4.4a,b) establish a unique reference space-time
that is neither boost
related nor rotating with respect to the physical space-time.
These conditions uniquely associate a set of tetrad fields to
an arbitrary metric tensor.
Therefore in the present framework it does not suffice to
assert that the reference space-time is Minkowski's space-time.
It is also necessary to enforce the soldering of the reference
space-time to the physical space-time by means of
Eqs. (4.4a,b). We conjecture that for a given space
volume the latter conditions yield the {\it minimum}
value for the energy expression (3.3).

\bigskip
\bigskip
\bigskip
\noindent {\it Acknowledgements}\par
\noindent T. M. L. T., J. F. R. N. and K. H. C. B.
are grateful to the Brazilian agency CAPES for financial
support.\par
\bigskip
\bigskip

\centerline{\bf APPENDIX}
\bigskip
We present here the components of the torsion tensor obtained out
of the tetrad configuration Eq. (4.7),
that satisfies M\o ller's weak field approximation:

\begin{eqnarray}
{T^{(0)}}_{01} & = & \sqrt{1+M^2y^2}\left(\fff{\psi}
{\rho^2}\partial_r\rho - \fff{1}{\rho} \partial_r\psi\right)
- \fff{\psi My}{\rho\sqrt{1+M^2y^2}}\left(y \partial_r M
+ M\partial_r y\right) ,
\nonumber
\\
{T^{(0)}}_{13} & =
& \fff{yN\chi}{\rho \psi}
\sen^2\theta \left(\fff{1}{y}\partial_r y
+ \fff{1}{\chi} \partial_r \chi + \fff{1}{N} \partial_r N -
\fff{1}{\rho} \partial_r \rho -
\fff{1}{\psi} \partial_r \psi\right) ,
\nonumber
\\
{T^{(1)}}_{01} & = &  -
\fff{y\chi}{\rho\Sigma}\sen\theta\sen\phi
\left(\fff{1}{\chi}\partial_r\chi + \fff{1}{y} \partial_r y
- \fff{1}{\rho} \partial_r\rho -
\fff{1}{\Sigma} \partial_r\Sigma\right) ,
\nonumber
\\
{T^{(1)}}_{03} & = & -\fff{y\chi}{\rho \Sigma}
\sen \theta \cos \phi ,
\nonumber
\\
{T^{(1)}}_{12} & = & \cos\theta\cos\phi \left(\partial_r\rho
-\fff{\rho}{\sqrt{\Delta}}\right) -
\fff{1}{\sqrt{\Delta}}\sen\theta\cos\phi\,\partial_\theta\rho,
\nonumber
\\
{T^{(1)}}_{13} & = &
\sen\theta\sen\phi\left[\fff{\rho}{\sqrt{\Delta}}-
\fff{\Sigma}{\rho}
\sqrt{1+N^2M^2y^2}\left(\fff{1}{\Sigma}\partial_r \Sigma
- \fff{1}{\rho}\partial_r\rho\right) - \right.
\nonumber
\nonumber
\\
&&\left. -
\fff{\Sigma N^2M^2y^2}{\rho\sqrt{1 + N^2M^2y^2}}
\left(\fff{1}{N}\partial_r N + \fff{1}{M} \partial_r M +
\fff{1}{y}\partial_r y\right)\right] ,
\nonumber
\\
{T^{(2)}}_{01} & = &
\fff{y\chi}{\rho\Sigma}\sen\theta\cos\phi \left(\fff{1}{\chi}
\partial_r \chi + \fff{1}{y} \partial_r y -
\fff{1}{\rho} \partial_r\rho -
\fff{1}{\Sigma} \partial_r \Sigma\right) ,
\nonumber
\\
{T^{(2)}}_{03} & =
& -\fff{y\chi}{\rho\Sigma} \sen\theta \sen \phi ,
\nonumber
\\
{T^{(2)}}_{12} & = &
\cos \theta \sen\phi \left(\partial_r\rho
- \fff{\rho}{\sqrt{\Delta}}\right) -
\fff{1}{\sqrt{\Delta}} \sen \theta
\sen \phi\, \partial_\theta \rho ,
\nonumber
\\
{T^{(2)}}_{13} & = &
-\sen\theta \cos\phi \left[\fff{\rho}{\sqrt{\Delta}}
- \fff{\Sigma}{\rho}
\sqrt{1+N^2M^2y^2} \left(\fff{1}{\Sigma}\partial_r \Sigma
- \fff{1}{\rho}\partial_r\rho\right) -\right.
\nonumber
\nonumber
\\
&&\left.- \fff{\Sigma}{\rho\sqrt{1 + N^2M^2 y^2}}
\left(\fff{1}{N}\partial_r N +
\fff{1}{M} \partial_r M +
\fff{1}{y} \partial_r y\right)\right] ,
\nonumber
\\
{T^{(3)}}_{12} & = &
-\sen\theta (\partial_r \rho - \fff{\rho}{\sqrt{\Delta}})
- \fff{1}{\sqrt{\Delta}} \cos \theta\, \partial_\theta \rho.
\nonumber
\end{eqnarray}

\newpage
\centerline{\bf FIGURE CAPTIONS}
\bigskip
\noindent {\bf Figure 1} - Energy within the external event
horizon of the Kerr black hole as a function of the angular
momentum. The figure displays
$\varepsilon = E/m$ against $\lambda$ for
expressions (5.3) and (5.4). The lower curve represents
$2M_{irr}$ given by Eq. (5.5). The one right above it, almost
coinciding with the lower curve, corresponds to Eq. (5.4).
The upper curve corresponds to Eq. (5.3).

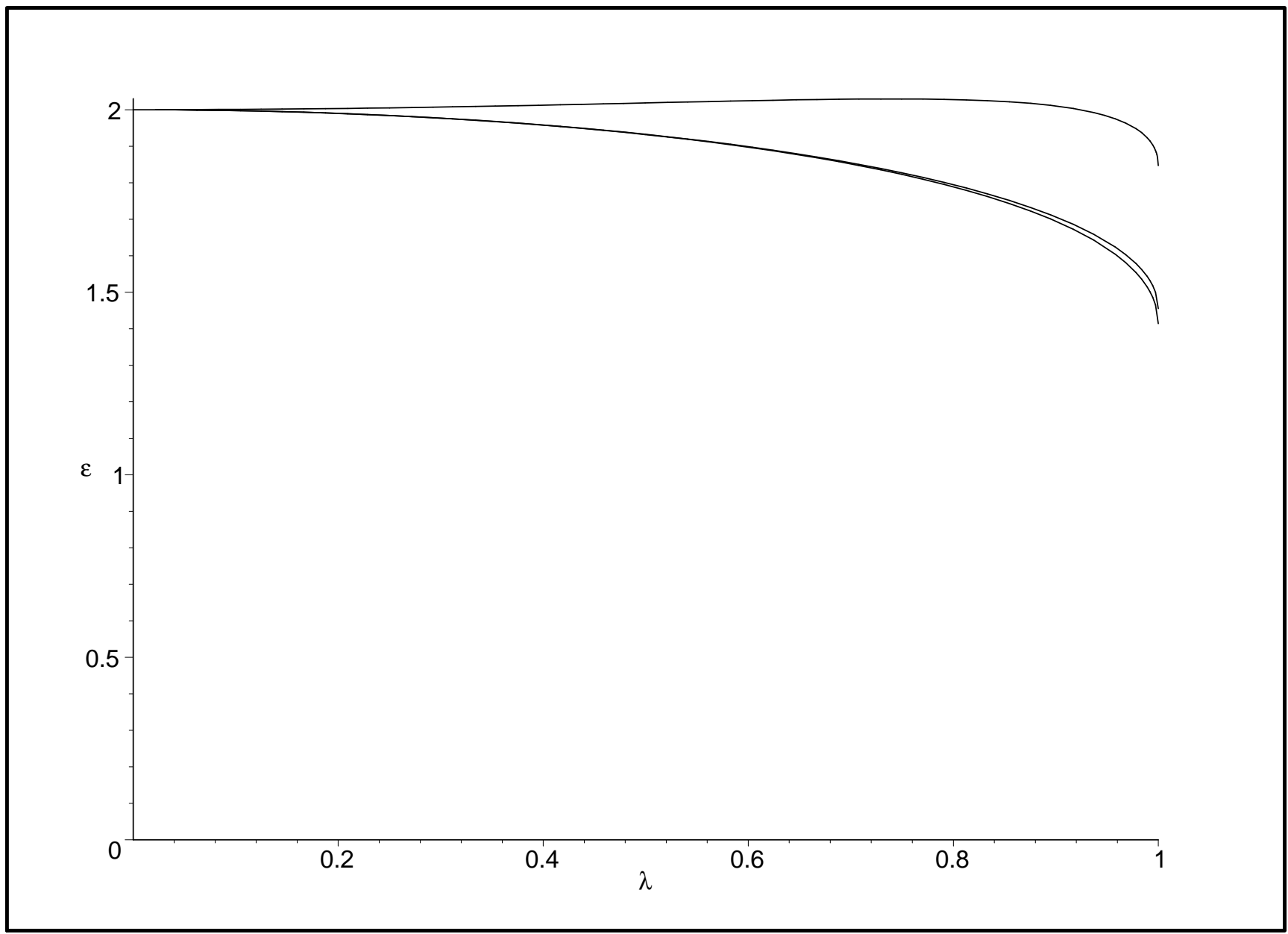
\end{document}